\documentclass[oneside]{elsart}

\usepackage{epsfig}
\usepackage{rotate}
\usepackage{amsmath}
\usepackage{amssymb}
\usepackage{array}
\usepackage[dvips]{color}
\usepackage{wrapfig}
\usepackage{bm}

\newcommand{\vo}{\mbox{$V^0$}}
\newcommand{\lam}{\mbox{$\rm \Lambda$}}
\newcommand{\lamdecay}{\mbox{$\rm \Lambda \to p \pi^-$}}
\newcommand{\alam}{\mbox{$\rm \bar \Lambda$}}
\newcommand{\alamdecay}{\mbox{$\rm \bar \Lambda \to \bar p \pi^+$}}
\newcommand{\ko}{\mbox{$\rm K^0_s$}}
\newcommand{\kodecay}{\mbox{$\rm K^0_s \to \pi^+ \pi^-$}}

\newcommand{\numuCC}{$\nu_\mu\ CC$}

\begin{document}
\begin{frontmatter}
\title{\boldmath 
A Study of Strange Particles Produced in \\ Neutrino 
Neutral Current Interactions \\ in the NOMAD Experiment}

\centerline{\bf NOMAD Collaboration}
\vskip 0.2cm
\author[Dubna]             {D.~Naumov\corauthref{corr}},
\corauth[corr]             {Corresponding author.}
\ead                       {naumov@nusun.jinr.ru}
\author[Dubna]             {A.~Chukanov},
\author[Dubna]             {E.~Naumova},
\author[Dubna,Paris]       {B.~Popov},
\author[Paris]             {P.~Astier},
\author[CERN]              {D.~Autiero},
\author[Saclay]            {A.~Baldisseri},
\author[Padova]            {M.~Baldo-Ceolin},
\author[Paris]             {M.~Banner},
\author[LAPP]              {G.~Bassompierre},
\author[Lausanne]          {K.~Benslama},
% ...

\address[LAPP]           {LAPP, Annecy, France}
% ...
\address[Dubna]          {JINR, Dubna, Russia}
% ...
\address[CERN]           {CERN, Geneva, Switzerland}
\address[Lausanne]       {University of Lausanne, Lausanne, Switzerland}
% ...
\address[Padova]         {Univ. of Padova and INFN, Padova, Italy}
\address[Paris]          {LPNHE, Univ. of Paris VI and VII, Paris, France}
% ...
\address[Saclay]         {DAPNIA, CEA Saclay, France}
% ...

\thanks[Scuola]          {Now at Scuola Normale Superiore, Pisa, Italy}
\thanks[Perugia]         {Now at Univ. of Perugia and INFN, Perugia, Italy}

\clearpage
\begin{abstract}
Results of a detailed study of strange particle production in neutrino 
neutral current interactions are presented using the data from the
NOMAD experiment. Integral yields of neutral strange particles 
($\ko$, $\lam$, $\alam$) have been measured.
Decays of resonances and heavy hyperons with an identified 
$\ko$ or $\lam$ in the final state have been analyzed. 
Clear signals corresponding to $\rm {K^\star}^\pm$ and $\rm {\Sigma(1385)}^\pm$
have been observed.
First results on the measurements of the $\lam$ 
polarization in neutral current interactions have been obtained.
\end{abstract}
\end{frontmatter}

\section{Introduction}

A study of strange particle production in neutrino neutral current
(NC) interactions is important, 
especially taking into account the fact
that strange particle production in $\nu$ NC is different from neutrino
charged current (CC) interactions 
at the quark level. For example, NC interactions can produce 
a leading down quark, a leading up quark and even a leading strange
quark, whereas CC interactions are a source of leading up and charm
quarks only.

A previous study of neutral strange particle production in $\nu_\mu$ NC 
interactions has been performed by the E632 Collaboration~\cite{E632} 
using the Fermilab 15-foot bubble chamber
exposed to the Tevatron (anti)neutrino beam with an average energy of
$\langle E_{\nu (\bar \nu)} \rangle$ = 150 (110) GeV. 
The selected $\vo$ sample in NC events is presented in Table~\ref{tab:E632},
and amounts to 81 events only. 
The $z = E_{V^0}/E_{had}$ distributions have also been  published~\cite{E632}.
\begin{table}[htb]
\begin{center} 
\caption{\label{tab:E632}%
\it 
  The number of reconstructed $\vo$ events and the
  corrected $\vo$ rates (yield per event) in $\nu_\mu$ NC interactions 
  for different cuts on the visible hadronic energy, $p_{had}^{vis}$,
  as measured in
  the E632 experiment (from~\cite{E632}). 
} 
\vspace*{0.3cm}
\begin{tabular}{ccc}
\hline
\hline
& $p_{had}^{vis} > 10 \ GeV$ & $p_{had}^{vis} > 25 \ GeV$ \\
\hline
\hline
NC events         & 670    & 437 \\
\hline
Observed $K^0$    & 48    & 28 \\
Observed $\lam$    & 29    & 19 \\
Observed $\alam$    & 4    & 2 \\
\hline
Corrected $K^0$ rate   & $0.344 \pm 0.065$  & $0.322 \pm 0.073$ \\
Corrected $\lam$ rate   & $0.111 \pm 0.025$  & $0.113 \pm 0.030$ \\
\hline
\hline
\end{tabular}
\end{center} 
\end{table}

In our recent articles~\cite{nomad-lambda-polar,nomad-alambda-polar}
we reported  results on the measurements of the polarization
of $\Lambda$ and $\bar \Lambda$ hyperons in $\nu_\mu$ CC
interactions. This is an
active field of research intimately related to the proton spin puzzle,
spin transfer mechanisms and fragmentation processes. Following our publication~\cite{nomad-lambda-polar}
the longitudinal $\Lambda$ polarization was investigated in much detail in a theoretical 
paper~\cite{EKN} in which agreement was found between the tuned polarized intrinsic model predictions and NOMAD
results. It is a natural step to confront the predictions of this model for the longitudinal polarization
of $\Lambda$ hyperons produced in neutrino nucleon neutral current deep inelastic scattering (DIS)
with experimental measurements. The large statistics of the NOMAD data (about 300K reconstructed NC events)
provides a tool to perform such a study.

\begin{figure}[h]
\begin{center}
   \mbox{
     \epsfig{file=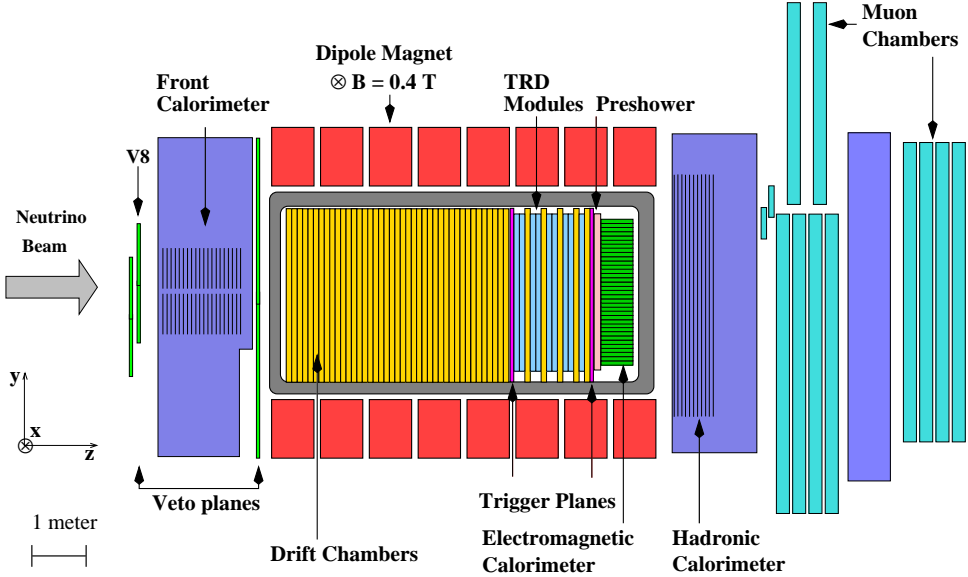,width=130mm}}
     \caption{A sideview of the NOMAD detector.}
      \label{fig:nomad_detector}
   \end{center}
\end{figure}

The NOMAD detector~\cite{NOMAD_DETECTOR} (see Fig.~\ref{fig:nomad_detector})
consisted of an active target of 44 drift chambers, with a total fiducial mass 
of 2.7~tons, located in a 0.4~Tesla dipole magnetic field. 
The coordinate system used in the NOMAD experiment is shown in Fig.~\ref{fig:nomad_detector}.
The drift chambers~\cite{NOMAD_NIM_DC}, made of low $Z$ material (mainly carbon)
served the double role of 
a (nearly isoscalar) target for neutrino interactions
and of the tracking medium. 
The average density of the drift chamber volume was 
0.1~$\mbox{g}/\mbox{cm}^3$.
These drift chambers provided 
an overall efficiency for charged track reconstruction 
of better than 95\% and a momentum 
resolution of approximately 3.5\% in the momentum range of interest
(less than 10~$\mbox{GeV}/\mbox{c}$).
Reconstructed 
tracks were used to determine the event topology 
(the assignment of tracks to vertices),
to reconstruct the vertex position and 
the track parameters at each vertex and, finally, to
identify the vertex type (primary, secondary, $\vo$, etc.).
The vertex position resolution is 600~$\mu$m, 90~$\mu$m and 860~$\mu$m
in $x$, $y$ and $z$ respectively. 
A transition radiation detector~\cite{NOMAD_TRD} 
placed at the end of the active target was used for electron 
identification.
A lead-glass electromagnetic calorimeter~\cite{NOMAD_ECAL} located
downstream of the tracking region provided 
an energy resolution of $3.2\%/\sqrt{E \mbox{[GeV]} } \oplus 1\%$
for electromagnetic showers and was essential to measure 
the total energy flow in neutrino interactions.
In addition, an iron absorber and a set of muon chambers located after 
the electromagnetic calorimeter were used for muon
identification, providing a muon detection 
efficiency of 97\% for momenta greater than 5~GeV/c.

The main characteristics of the neutrino beam~\cite{NOMAD_FLUX} are given in 
Table~\ref{tab:beam_info}.
\begin{table}[htb]
\begin{center}
\caption{\it The CERN SPS neutrino beam composition~\cite{NOMAD_FLUX}.}
\vspace*{0.3cm}
\begin{tabular}{||c|c|c|c|c||}
\hline
\hline
\multicolumn{1}{||c|}{Neutrino} & \multicolumn{2}{c}{Flux} 
& \multicolumn{2}{|c||}{CC interactions in NOMAD} \\
\cline{2-5}
flavours & $< E_\nu >$ [GeV] & rel.abund. & $< E_\nu >$ [GeV] & rel.abund. \\
\hline
\hline
$\nu_\mu$         & 24.3 & 1    & 47.5 &  1 \\
$\bar{\nu}_\mu$ &  17.2 &  0.0678  &  42.0 &  0.024 \\
$\nu_e$          &  36.4 &  0.0102  &  58.2 &  0.015 \\
$\bar{\nu}_e$  & 27.6 &  0.0027 & 50.9 &  0.0015 \\
\hline
\hline
\end{tabular}
\label{tab:beam_info}
\end{center}
\end{table}

Whenever possible the NOMAD data are compared to the results
of a Monte Carlo (MC) simulation based on LEPTO 6.1~\cite{LEPTO} and
JETSET 7.4~\cite{JETSET} generators for neutrino interactions and 
on a GEANT~\cite{GEANT} based program for the detector response. 
The relevant JETSET parameters have been tuned in order to reproduce
the yields of all hadrons, including strange particles, as
measured in $\nu_\mu$ CC interactions in NOMAD~\cite{nomad-strange-cc}. A detailed
paper devoted to the NOMAD MC tuning will be the subject of a forthcoming publication.
To define the parton content of the nucleon for the cross-section calculation
we have used the parton density functions parametrized in~\cite{Alekhin}.

This article is organized as follows. The event selection criteria
and a special algorithm developed for
the identification of neutrino NC interactions are described in Sec.~\ref{sec:nc-ident}. The production
yields of neutral strange particles ($\ko$, $\lam$ and $\alam$) are presented in 
Sec.~\ref{sec:v0_yields}, while yields of heavier resonances 
($\rm {K^\star}^\pm$ and $\rm {\Sigma(1385)}^\pm$) are reported in Sec.~\ref{sec:resonances}. A measurement
of the $\lam$ polarization vector is presented in Sec.~\ref{sec:polar}. Finally, we summarize our
results, compare them to the measurements in the
\numuCC \ event sample and present our conclusions in Sec.~\ref{sec:conclusions}.

\section{\label{sec:nc-ident}Identification of neutrino NC interactions}

In this 
section we describe the selection procedure for 
NC DIS events. An event with no {\em identified muon} is considered
as a candidate for a NC interaction. However, such a simplified
identification would lead to a
considerable contamination from neutrino charged current 
events with {\em a reconstructed} muon which was {\em not identified}
as a muon by the reconstruction algorithm based on the
information from the muon chambers.
We estimate such 
a contribution from $\nu_\mu$ CC events with the help of the
MC simulation to be about 
30\% of NC candidates. 
However, the difference in kinematic properties of NC and CC events can be used to 
discriminate between these two event categories. 
For this purpose we apply an identification of NC events which
was used in~\cite{oscill} and consists of two main steps:
\begin{enumerate}
\item Muon tagging, finding the negatively charged track most likely to be a muon,
based on the topology of the event.
\item CC rejection, using the likelihood based on event kinematics that the candidate muon
originates indeed from a $\nu_\mu$ CC interaction.
\end{enumerate}
To tune our identification algorithm we use two MC subsamples 
(CC and NC) each corresponding to one half of the total MC statistics (well over $10^6$ events)
and we require that:
\begin{enumerate}
\item There is no identified negatively charged muon in the event.
\item The primary vertex of the event is in the restricted fiducial volume:\\
$|X,Y| < 120 \mbox{ cm}$, $5 \mbox{ cm} < Z < 350 \mbox{ cm}$.
\item There should be at least two charged tracks at the primary vertex in the event
of which at least one must be negative.
\item The visible hadronic energy should be larger than 3 GeV.
\end{enumerate}
The actual NC selection efficiency and background contamination are then determined using
the second halves of the MC event samples.

\subsection{Muon Tagging\label{sec:muon_tag}}
\begin{figure}[htb]
\centering\epsfig{file=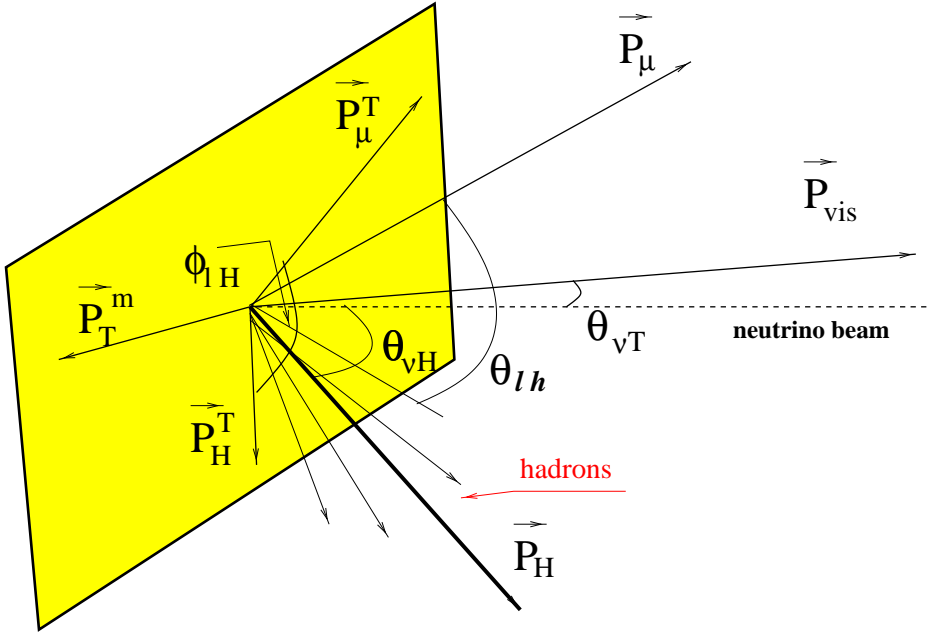,width=0.9\textwidth} 
\caption{Definition of kinematic variables used for the event tagging. 
The incoming neutrino direction is along the $Z$ axis shown by 
the dashed line.  \label{fig:varnc} }
\end{figure}

\begin{figure}[htb]
\epsfig{file=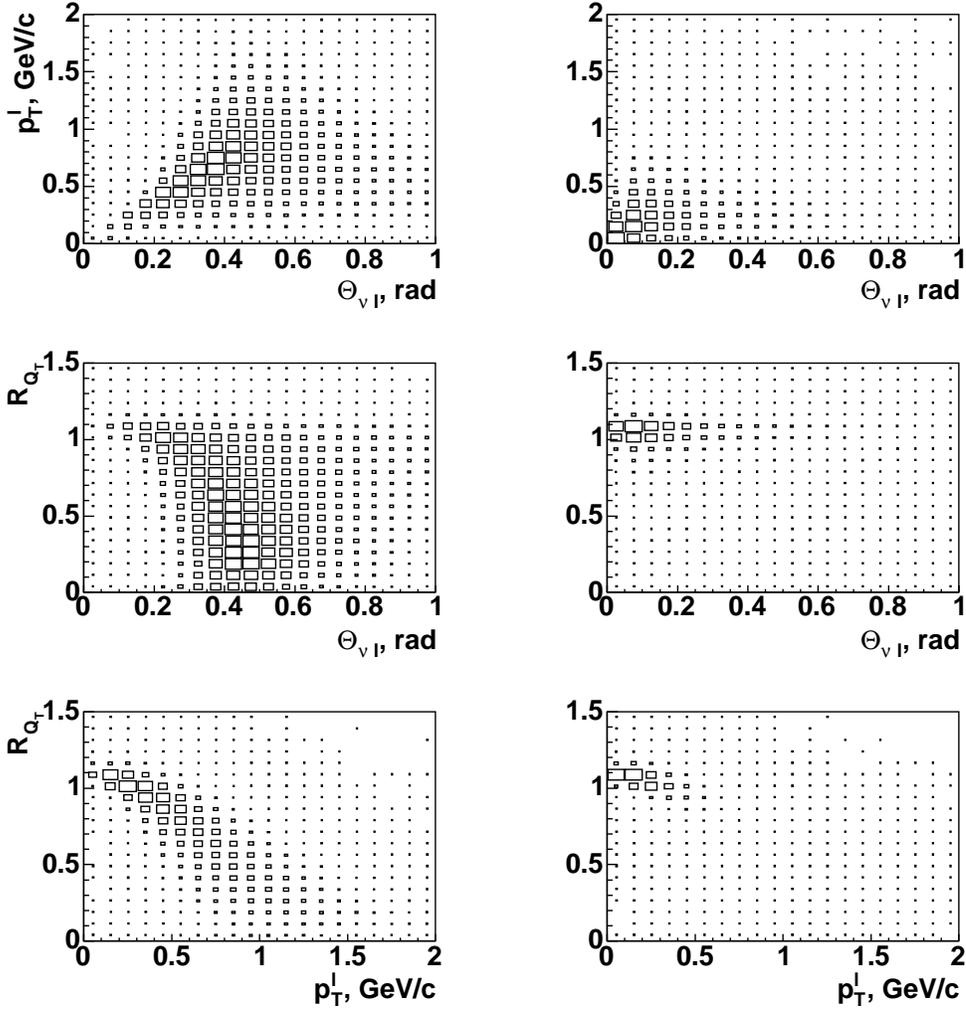,width=\textwidth} 
\caption{\label{fig:muon_vars_cc} Correlations between the kinematic
variables $\theta_\nu^l$, $P_T^l$ and $R_Q$ for reconstructed but not identified muons 
(left plots) and for negatively charged primary hadrons (right plots) in 
$\nu_\mu$ CC events.}
\end{figure}

In what follows we describe our procedure to tag the primary muon produced at the charged
current vertex. The kinematic variables used in this procedure are defined in Fig.~\ref{fig:varnc}.
We have found that the following three variables 
provide the best discrimination 
between the true primary $\mu^-$ and any other negatively charged track from the hadronic jet:
\begin{enumerate}
\item $\theta_{\nu l}$ - the angle between the track and the incoming 
neutrino direction (Z axis),
\item $p_{T}^{l}$ - the transverse momentum of the track with respect to the incoming 
neutrino direction,
\item $R_{Q_{T}} = \langle Q_T^2 \rangle_{H}/\langle Q_T^2 \rangle _T$, where: 
\begin{eqnarray*}
&&\langle Q_T^2 \rangle_{H} = \frac{1}{n-1} \sum_{j\ne i} Q_T^2(j),\\
&&\langle Q_T^2 \rangle _T= \frac{1}{n} \sum_{j} Q_T^2(j),\\
&&\mathbf{Q_T} = \mathbf{p}_i - \mathbf{p}_{\rm vis} 
\frac{\mathbf{p}_i\cdot\mathbf{p}_{\rm vis}}{\mathbf{p}_{\rm vis}^2}, 
\end{eqnarray*}
$\mathbf{p}_i$ is the momentum of the $i$-th track (assumed to be the candidate muon), and
$\mathbf{p}_{\rm vis}$ is the total momentum of the event, 
and $n$ is the total number of tracks in the event.
\end{enumerate}
As one can see from Fig.~\ref{fig:muon_vars_cc}
the distributions of these variables in CC events with an unidentified muon are quite different
if the $i$-th track is the true muon track or, instead, another negative primary track belonging to the
hadronic system. The tagging algorithm is based on the following three-dimensional likelihood function:
\begin{equation}
\label{eq:tag_muon}
{\mathcal{L}}^{\mu} = [\theta_{\nu l},p_{T}^{l},R_{Q_{T}}] 
\end{equation}
where the square brackets denote the correlations among the variables.
A likelihood ratio, $\ln \lambda^\mu$, is built from $\nu_\mu$ CC events with unidentified muons
as the ratio of ${\mathcal{L}}^{\mu}$ for the true muon and for other negative tracks. We tag as
candidate muon the primary negative track with the largest value of $\ln \lambda^\mu$.  
This procedure correctly identifies 86\%
of $\mu^-$ in $\nu_\mu$ CC events with no identified muon, and 71\% of $e^-$ in $\nu_e$ CC
events as well. Thus the muon tagging procedure can also be applied  
for the kinematical identification of primary electrons from
$\nu_e$ CC events.

\subsection{CC Event Rejection \label{sec:likelihoods}}

The requirement of a tagged muon to be a track with maximum $\ln \lambda^\mu$ 
always provides us with a muon candidate, even in the case of NC events. 
However, it is possible to discriminate between 
NC and CC events using a multi-dimensional likelihood function which exploits the
full event kinematics:
\begin{equation}
\label{eq:like6}
\mathcal{L}^{\rm NC}_{6}
= \left[\left[\left[\theta_{\nu H},\theta_{\nu T}\right],\theta_{l h_{i}},Q_T\right],{p_T^m},\phi_{l H}\right]
\end{equation}
where:
\begin{itemize}
\item $\theta_{\nu H}$ is the angle between the momentum of the hadronic system and the incident neutrino
direction;
\item $\theta_{\nu T}$ is the angle between the total visible momentum and the incident neutrino direction;
\item $\theta_{l h_{i}}$ is the minimum opening angle between any primary track and the muon candidate;
\item $p_T^m$ is the missing transverse momentum;
\item $\phi_{l H}$ is the azimuthal angle between the muon candidate and the hadronic system.
\end{itemize}

\begin{figure}[htb]
\centering\epsfig{file=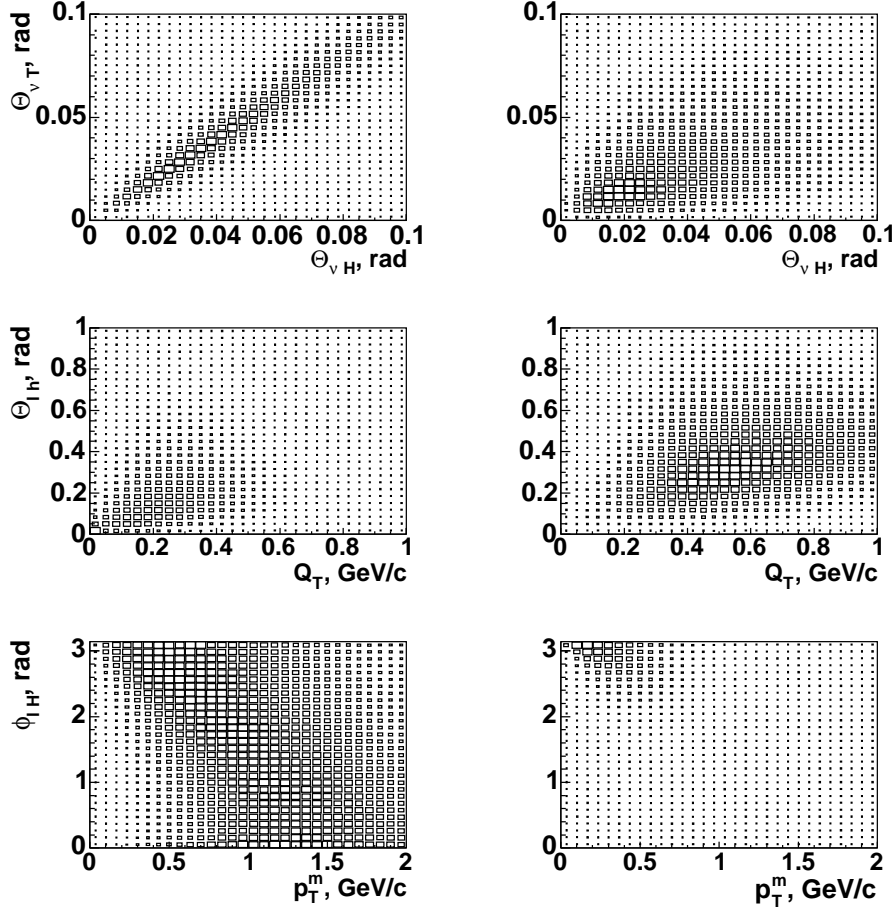,width=0.9\textwidth} 
\caption{\label{fig:likenc6d_vars_cc_2} Correlations  
among kinematic variables used to construct ${\mathcal{L}^{\rm NC}_6}$
for negatively charged tracks with the maximal $\ln \lambda^\mu$ 
in  $\nu_\mu$ NC events (left plots)  and for true muons in $\nu_\mu$ CC events 
(right plots).}
\end{figure}

Some of the correlations among the kinematic variables used to build $\mathcal{L}^{\rm NC}_{6}$
are shown in Fig.~\ref{fig:likenc6d_vars_cc_2} for true muons in $\nu_{\mu}$ CC events with unidentified
muons, and for the primary negative track with the largest value of $\ln \lambda^\mu$ in NC events.
A likelihood ratio, $\ln \lambda^{\rm NC}_{6}$, is built
as the ratio of the ${\mathcal{L}}^{\rm NC}_6$ function for 
the negative track tagged as the muon candidate in NC events, and for
the true muon in $\nu_{\mu}$ CC events with
unidentified muons.
All events with $\ln \lambda^{\rm NC}_{6} < 0.5$ are classified as CC interactions and rejected. 

As a cross-check, we also use a different likelihood which depends on three variables only:

\begin{equation}
\label{eq:like3}
\mathcal{L}^{\rm NC}_3 = [{p_T}^l,\rho,R_{Q_{T}}]
\end{equation}
where $$\rho = \frac{{p_T}^l}{{p_T}^l + {p_T}^m}.$$

\subsection{Efficiency and Purity\label{sec:eff_purity_NC}}

The NC identification procedure applied to both real and simulated events
greatly reduces the contamination of charged current events with leading
leptons not identified by the corresponding subdetectors, while keeping more than 60\% of the 
useful signal.

In order to estimate the final purity of the $\nu$ NC sample in the 
data we processed in addition small samples of $\bar\nu_\mu$ CC, $\nu_e$ CC and $\bar\nu_e$ CC 
events through the same code. The final $\nu$ NC purity is computed as:
\begin{equation*}
\label{eq:purity_beam}
Purity = \frac{\sum_\alpha f_{\alpha}\epsilon(\alpha NC \to NC)}
{\sum_{\alpha=\nu_\mu,\bar\nu_\mu, \nu_e,\bar\nu_e} 
f_{\alpha}\left(\frac{\sigma_{\alpha CC}}{\sigma_{\alpha NC}} \epsilon(\alpha CC\to NC)+
\epsilon(\alpha NC\to NC)\right)},
\end{equation*}
where $f_{\alpha}$ describe the beam composition (presented in
Table~\ref{tab:beam_info}) of $\bar\nu_\mu$, $\nu_e$ and $\bar\nu_e$ relative to the $\nu_\mu$ component,
$\sigma_{\alpha CC}$ and $\sigma_{\alpha NC}$ stand for the corresponding CC and NC cross-sections, and $\epsilon(\alpha CC(NC) \to NC)$ is the efficiency for the beam component $\alpha$
interacting via CC (NC) to be identified as NC signal. 
We assume that $\epsilon(\alpha NC \to NC)$ are all equal. 
\begin{figure}[htb]
  \centering\epsfig{file=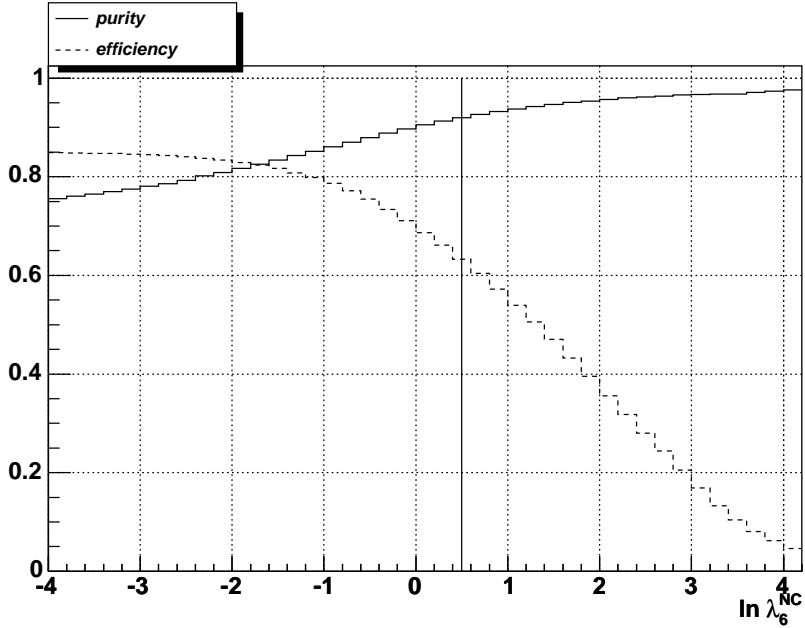,width=0.9\textwidth} 
  \caption{\label{fig:eff_pur_6} Efficiency and purity of the NC
  identification as a function of a cut on the $\ln \lambda^{\rm NC}_{6}$. The final
  cut used is also shown.}
\end{figure}
The final purity and efficiency of the NC sample identified with $\ln \lambda^{\rm NC}_{6}$ 
are displayed in Fig.~\ref{fig:eff_pur_6} as a function of the cut on $\ln \lambda^{\rm NC}_{6}$.
The use of $\ln \lambda^{\rm NC}_{3}$ as a cross-check gives a consistent number of
NC events identified in the data, thus confirming our NC identification procedure. 

In what follows we present our results with the $\ln \lambda^{\rm NC}_{6}$ identification, 
since for a given $\nu$ NC efficiency $\ln \lambda^{\rm NC}_{6}$ provides
a better rejection of $\bar\nu_\mu$, $\nu_e$ and $\bar\nu_e$ CC events.

A sample of 226681 $\nu$ NC events is selected in the data
with purity of 92\% using the cut $\ln \lambda_{6}^{\rm NC}>0.5$. 
The $\nu_\mu$ NC events being the dominating 
component of the final $\nu$ NC sample is selected with 62.7\% efficiency.

\begin{figure}[htb]
  \centering\epsfig{file=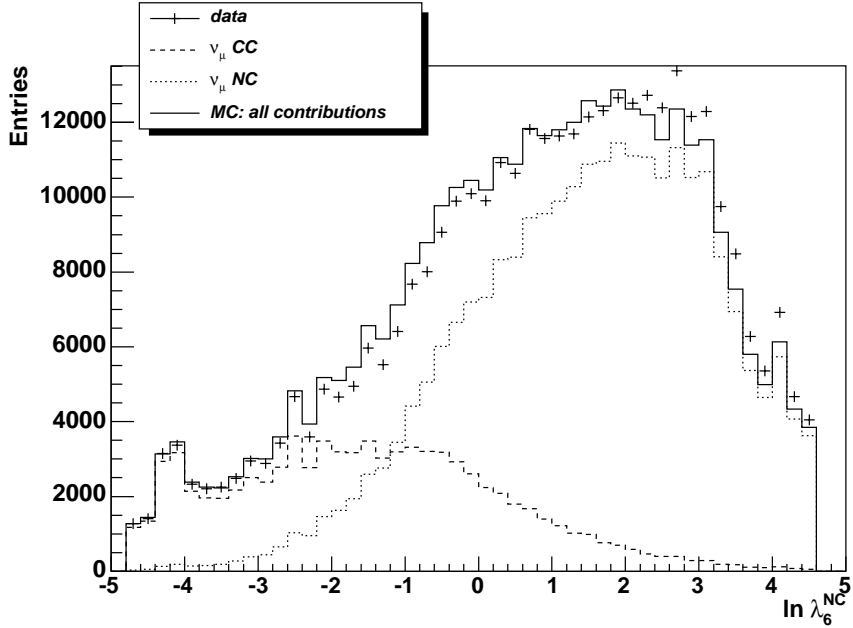,width=0.9\textwidth} 
  \caption{\label{fig:likelihood_compare} Likelihood ratio $\ln \lambda^{\rm NC}_{6}$ distributions in the
  data (points) and in the MC (histogram) normalized to the number of events in the data 
taking into account the actual beam composition. The main contributions from $\nu_\mu$ CC 
and $\nu_\mu$ NC simulated events are shown separately.}
\end{figure}

Fig.~\ref{fig:likelihood_compare} shows a comparison between the data and
MC (normalized to the number of events in the data taking into account the
actual beam composition) as a function of the $\ln \lambda^{\rm NC}_{6}$.
These distributions can be seen to be in good agreement.

\section{\label{sec:v0_yields}Production of neutral strange particles}

We apply the $V^0$ identification algorithm developed for the CC event sample in 
\cite{nomad-lambda-polar} to identify  neutral strange particles ($\lam$, $\ko$ and $\alam$) 
in $\nu_\mu$ NC sample. We
find values for efficiencies and purities of the $\vo$ identification algorithm
similar to those observed in the CC sample. 

The numbers of neutral strange particles observed in the data 
with their identification purities are summarized in Table~\ref{tab:v0stat}.
This is a considerable improvement in statistics over the previously published data~\cite{E632}.
\begin{table}[htb]
\caption{ \it Number of  $\vo$'s in the data and their identification purities 
in NC interactions \label{tab:v0stat}
}
\vspace*{0.2cm}
\begin{center}
\begin{tabular*}{0.4\linewidth}{@{\extracolsep{\fill}}ccc}
\hline
\hline
\# & DATA & Purity\\
\hline
$N(\raisebox{0.ex}[3ex][2ex]{\ko)}$   & 3691 & 97\% \\
$N(\raisebox{0.ex}[3ex][2ex]{\lam)}$  & 1619 & 94\% \\
$N(\raisebox{0.ex}[3ex][2ex]{\alam)}$ & 146  & 82\% \\
\hline
\hline
\end{tabular*}
\end{center}
\end{table}

\begin{figure}[htb]
\begin{tabular}{cc}
\epsfig{file=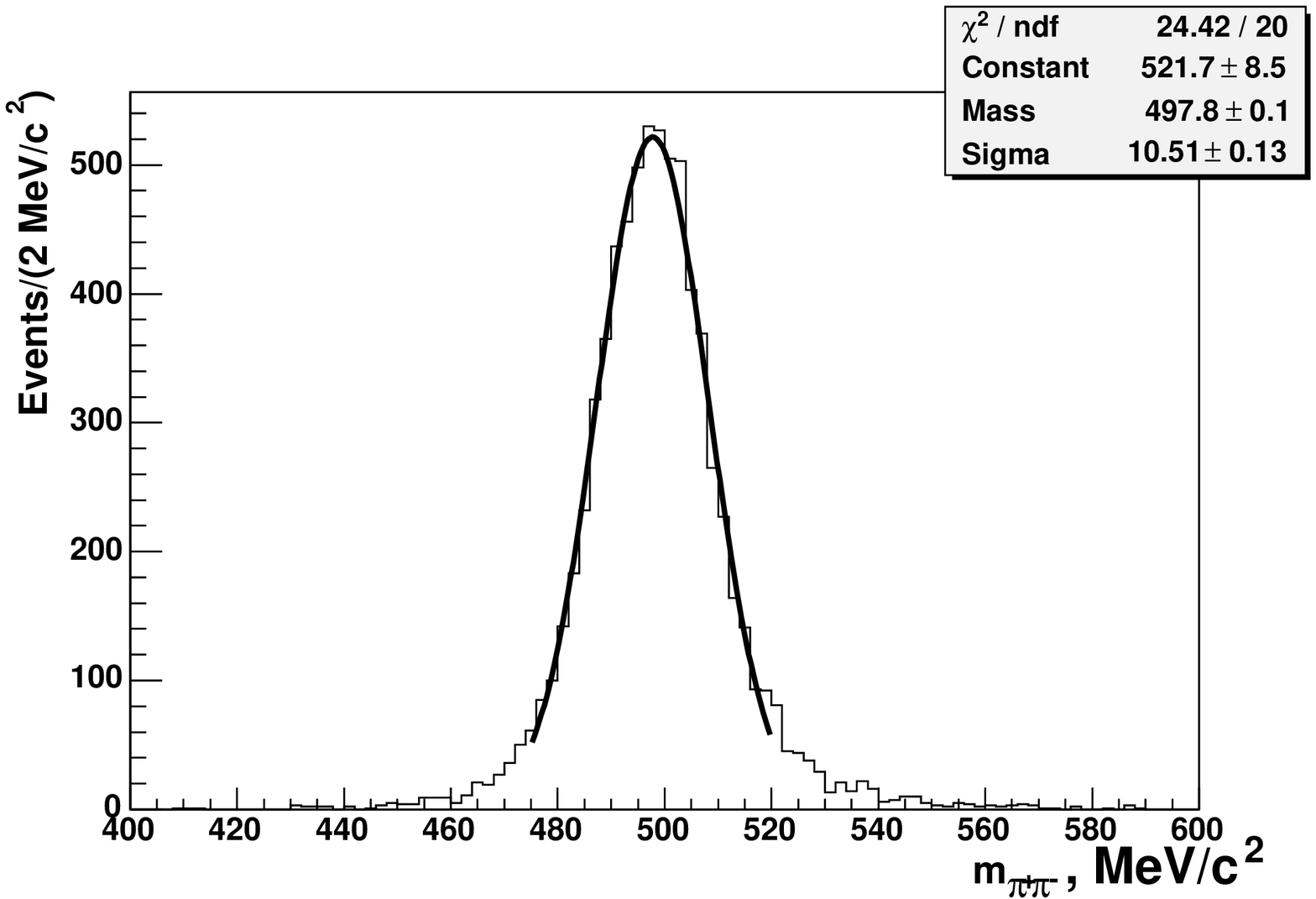,width=0.51\textwidth} &
\epsfig{file=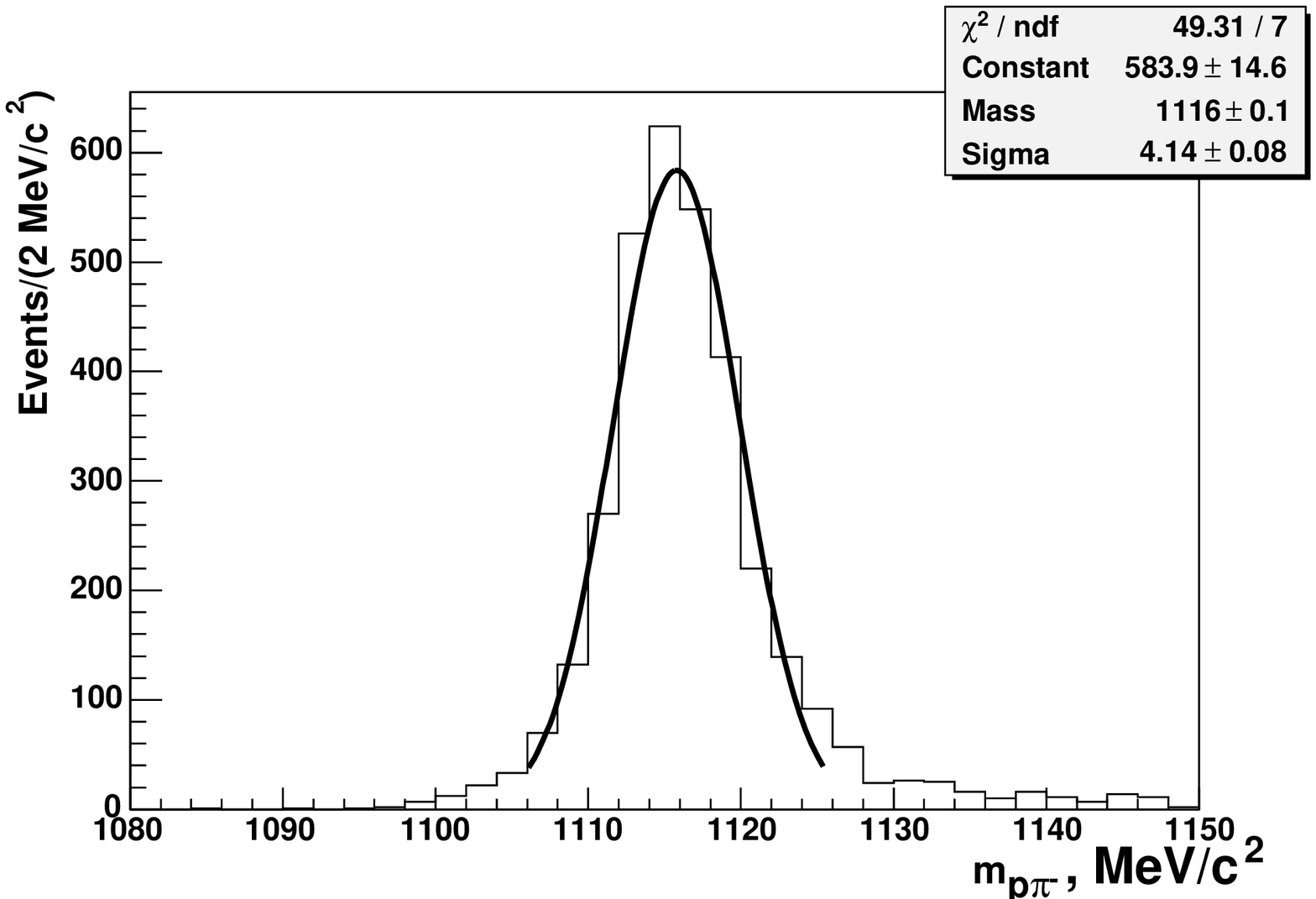,width=0.51\textwidth} \\
\end{tabular}
\begin{picture}(0,0)
\put(120,100){{\Large \bf $\ko$}}
\put(340,100){{\Large \bf $\lam$}}
\end{picture}
\caption{\label{fig:v0_masses} Invariant mass distributions of two charged tracks from a 
$\vo$ vertex identified as $\ko$ (left) and $\lam$ (right) in $\nu$ NC samples in data.}
\end{figure}

Fig.~\ref{fig:v0_masses} shows the invariant mass distributions
of two charged tracks from a $\vo$ vertex identified as $\ko$ and $\lam$ in the data. 
The mean values of these distributions agree well with the PDG world averages~\cite{PDG} for 
$\ko$ and $\lam$ masses respectively. 
The corresponding experimental resolutions are about 11 MeV$/c^2$ for $\kodecay$, and 
4 MeV$/c^2$ for both $\lamdecay$ and $\alamdecay$.

The integral yields of neutral strange particles per neutrino interaction have been measured
using the same approach as the one developed for the CC event sample~\cite{nomad-strange-cc}.
The results are presented in Table~\ref{tab:integral_yields}. 
We find agreement at the level of 12\% or better between data and the MC tuned on strange particle
production in CC events. 
This has to be compared with the factor of two overestimation of strange particle yields
in NC events by the MC with the default JETSET parameters~\cite{JETSET}.
\begin{table}[htbp]
\begin{center}
\caption{ \it Integral yields of $\vo$'s in NC interactions in both 
the data and MC simulation. Only statistical errors are shown.}
\vspace*{0.3cm}
\label{tab:integral_yields}
\begin{tabular*}{0.8\linewidth}{@{\extracolsep{\fill}}cccccc}
\hline
\hline
\vo \ Type & $\mathcal{T}\rm_{V^0}$ DATA & $\mathcal{T}\rm_{V^0}$ MC &
$\mathcal{T}\rm_{V^0}^{MC}/\mathcal{T}\rm_{V^0}^{DATA}$\\
\hline
\raisebox{0.ex}[3ex][2ex]{\lam}&$(5.16 \pm 0.14)\%$ & $(5.79 \pm 0.02) \%$ &$1.12 \pm 0.03$ \\
\raisebox{0.ex}[3ex][2ex]{\ko}& $(8.62 \pm 0.15)\%$ & $(8.46 \pm 0.03)\%$ & $0.98 \pm 0.02$\\
\raisebox{0.ex}[3ex][2ex]{\alam}& $(0.43 \pm 0.04)\%$ & $(0.44 \pm 0.01)\%$ & $1.02 \pm 0.11$ \\
\hline
\hline
\end{tabular*}
\end{center}
\end{table}

It is interesting to compare the measured yields of neutral strange particles in  
NC interactions with those in $\nu_\mu$ CC events. For such a comparison we 
applied the same kinematic criteria to CC events (except for the $Q^2>0.8$ GeV$^2$
cut applied for the CC sample and not applied for NC events) 
and obtained the yields summarized in Table~\ref{tab:integral_yields_cc}. We  emphasize
the excellent agreement of production yields of strange particles between the data
and MC CC samples obtained as a result of the MC tuning.
From Tables~\ref{tab:integral_yields} and~\ref{tab:integral_yields_cc} we conclude that 
the production yields of neutral strange particles in CC and NC events agree within a few percent.

We find a consistency between neutral strange particles production yields measured with our
default cut $\ln \lambda^{\rm NC}_{6}>0.5$ and with the harder cut $\ln \lambda^{\rm NC}_{6}>3$.
Varying the $V^0$ selection criteria~\cite{nomad-lambda-polar} as well
as the NC identification cut within reasonable limits we have
estimated the systematic uncertainties on the $V^0$ production yields
in neutrino NC interactions: 0.09 for $\lam$, 0.11 for $\ko$ and 0.03 for $\alam$.

\begin{table}[htbp]
\begin{center}
\caption{ \it Integral yields of $\vo$'s in $\nu_\mu$ CC interactions in both 
the data and MC simulation. Only statistical errors are shown.}
\vspace*{0.3cm}
\label{tab:integral_yields_cc}
\begin{tabular*}{0.8\linewidth}{@{\extracolsep{\fill}}cccccc}
\hline
\hline
\vo \ Type & $\mathcal{T}\rm_{V^0}$ DATA & $\mathcal{T}\rm_{V^0}$ MC &
$\mathcal{T}\rm_{V^0}^{MC}/\mathcal{T}\rm_{V^0}^{DATA}$\\
\hline
\raisebox{0.ex}[3ex][2ex]{\lam}&$(5.55\pm 0.08)\%$ & $(5.52\pm 0.02) \%$ &$0.99 \pm 0.02$ \\
\raisebox{0.ex}[3ex][2ex]{\ko}& $(8.54\pm 0.08)\%$ & $(8.51\pm 0.02)\%$ & $1.00 \pm 0.01$\\
\raisebox{0.ex}[3ex][2ex]{\alam}& $(0.43\pm 0.02)\%$ & $(0.46\pm 0.01)\%$ & $1.05 \pm 0.06$ \\
\hline\hline
\end{tabular*}
\end{center}
\end{table}

\begin{figure}[htb]
\center{%
\mbox{\epsfig{file=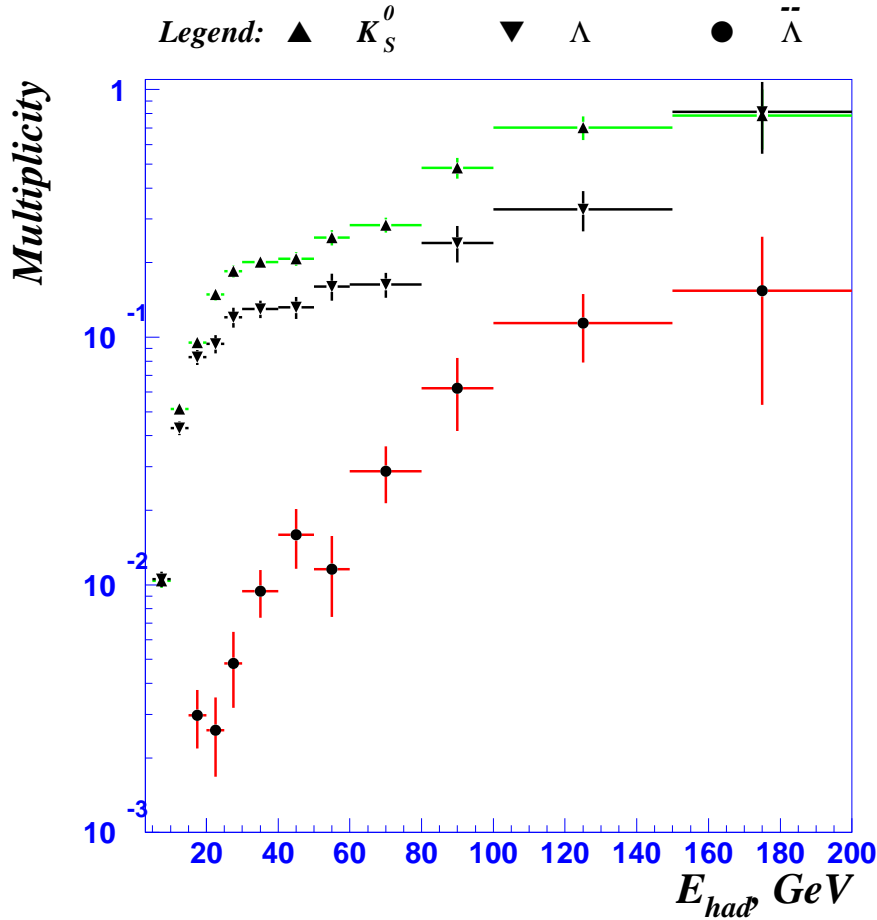,width=0.9\linewidth}}}
\protect\caption{\it
Dependence of the yields of $\ko$'s, $\lam$'s and $\alam$'s 
on the total hadronic energy in the NC data sample.
}
\label{fig:v0_yields}
\end{figure}

The dependence of the yields of $\ko$'s, $\lam$'s and $\alam$'s in
events in the data 
on the total hadronic energy $E_{had}$ is shown in
Fig.~\ref{fig:v0_yields}. As expected, the yields increase as a
function of $E_{had}$. 

We notice a reasonable agreement between NOMAD and E632~\cite{E632} 
measurements of neutral strange particles production rates. 

\begin{figure}[htb]
  \centering\epsfig{file=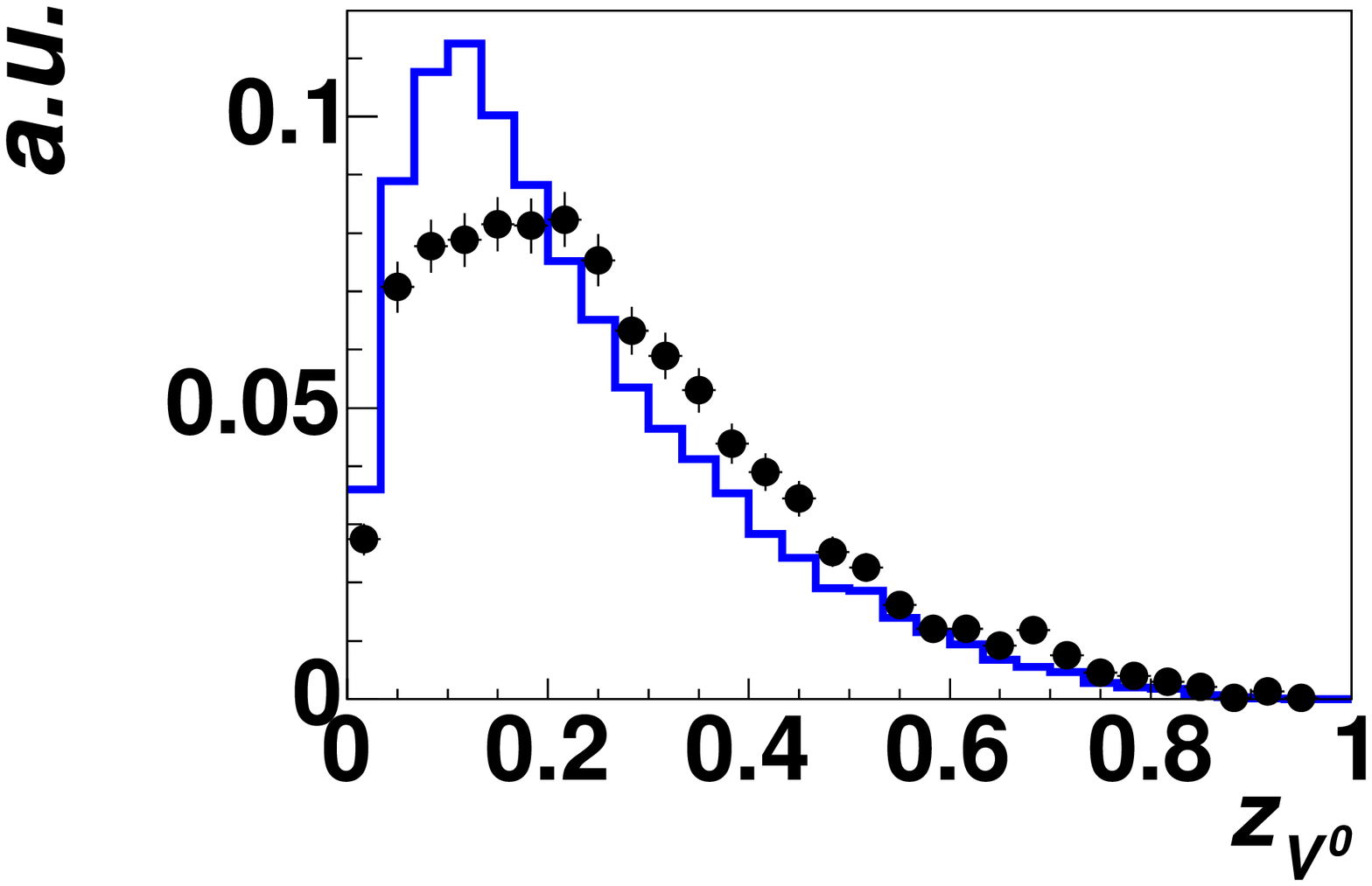,width=0.32\textwidth}
  \centering\epsfig{file=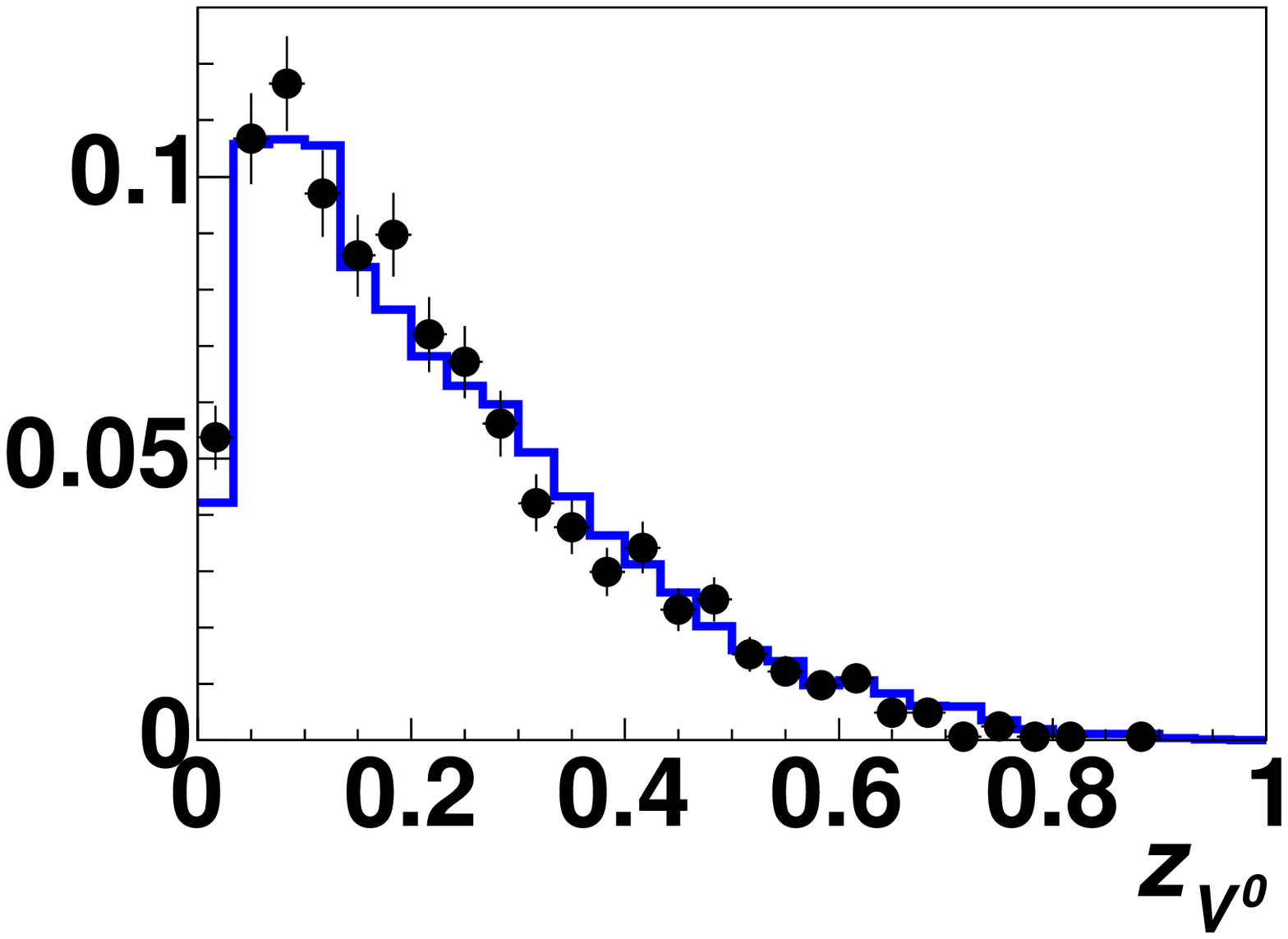,width=0.32\textwidth}
  \centering\epsfig{file=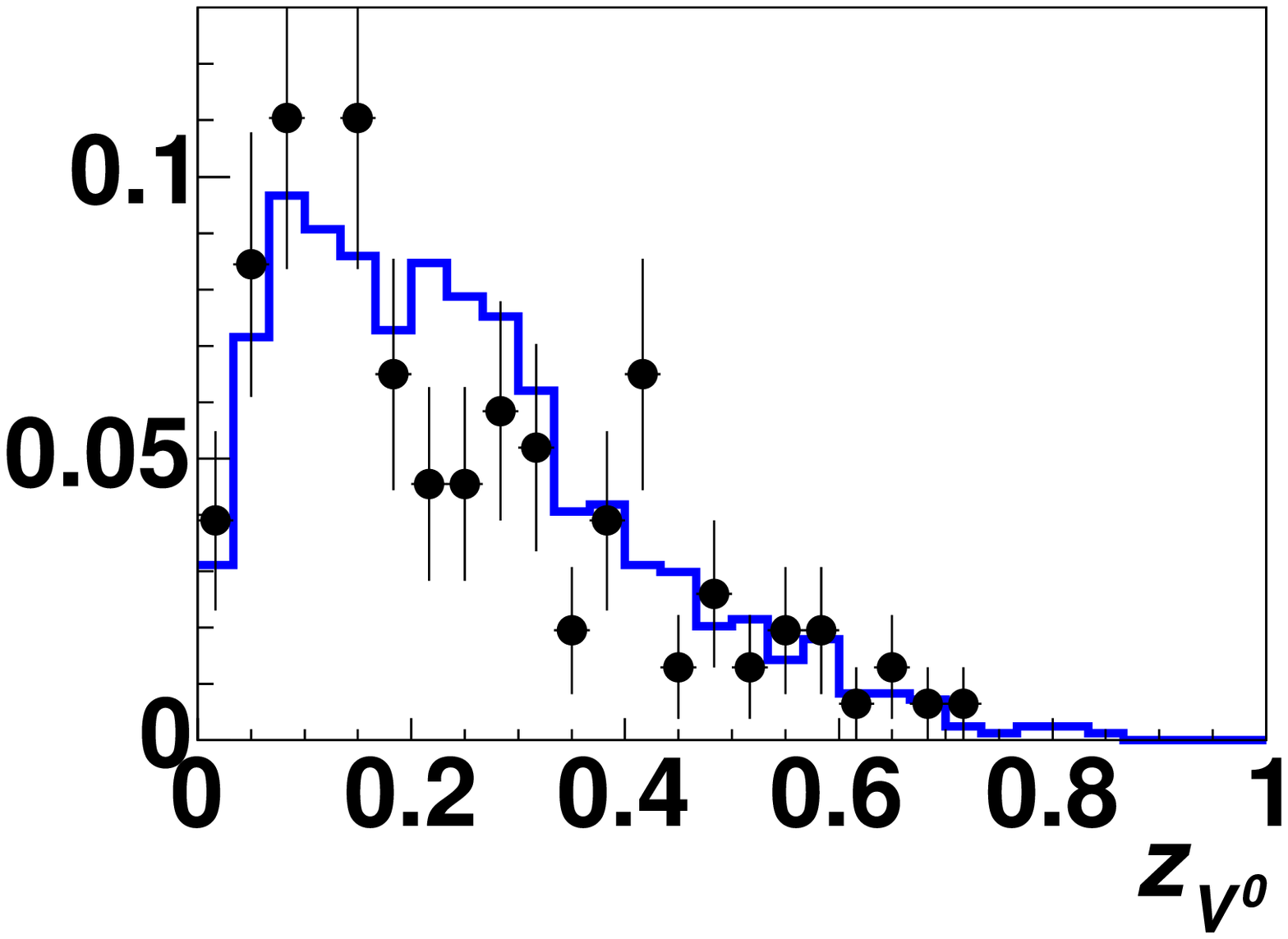,width=0.32\textwidth}
\begin{picture}(0,0)
\put(-320,50){{\Large \bf $\ko$}}
\put(-190,50){{\Large \bf $\lam$}}
\put(-50,50){{\Large \bf $\alam$}}
\end{picture}
  \caption{\label{fig:v0_injet} Reconstructed $z$-distributions for identified $\ko$'s
  (left plot), $\lam$'s (middle plot) and $\alam$'s (right plot)
  in $\nu$ NC samples both in the data (points with error bars) and in
  the MC (histogram).
}
\end{figure}

The distributions of the reconstructed $z = E_{V^0}/E_{had}$ variable for identified $\ko$'s
(left plot), $\lam$'s (middle plot) and $\alam$'s (right plot) 
in NC events are shown in Fig.~\ref{fig:v0_injet} for both data and MC samples.
There is an agreement between data and MC except for the $\ko$ case.

\section{\label{sec:resonances}Study of heavier strange particles and resonances}

\begin{figure}[htb]
\center{%
\begin{tabular}{cc}
\mbox{\epsfig{file=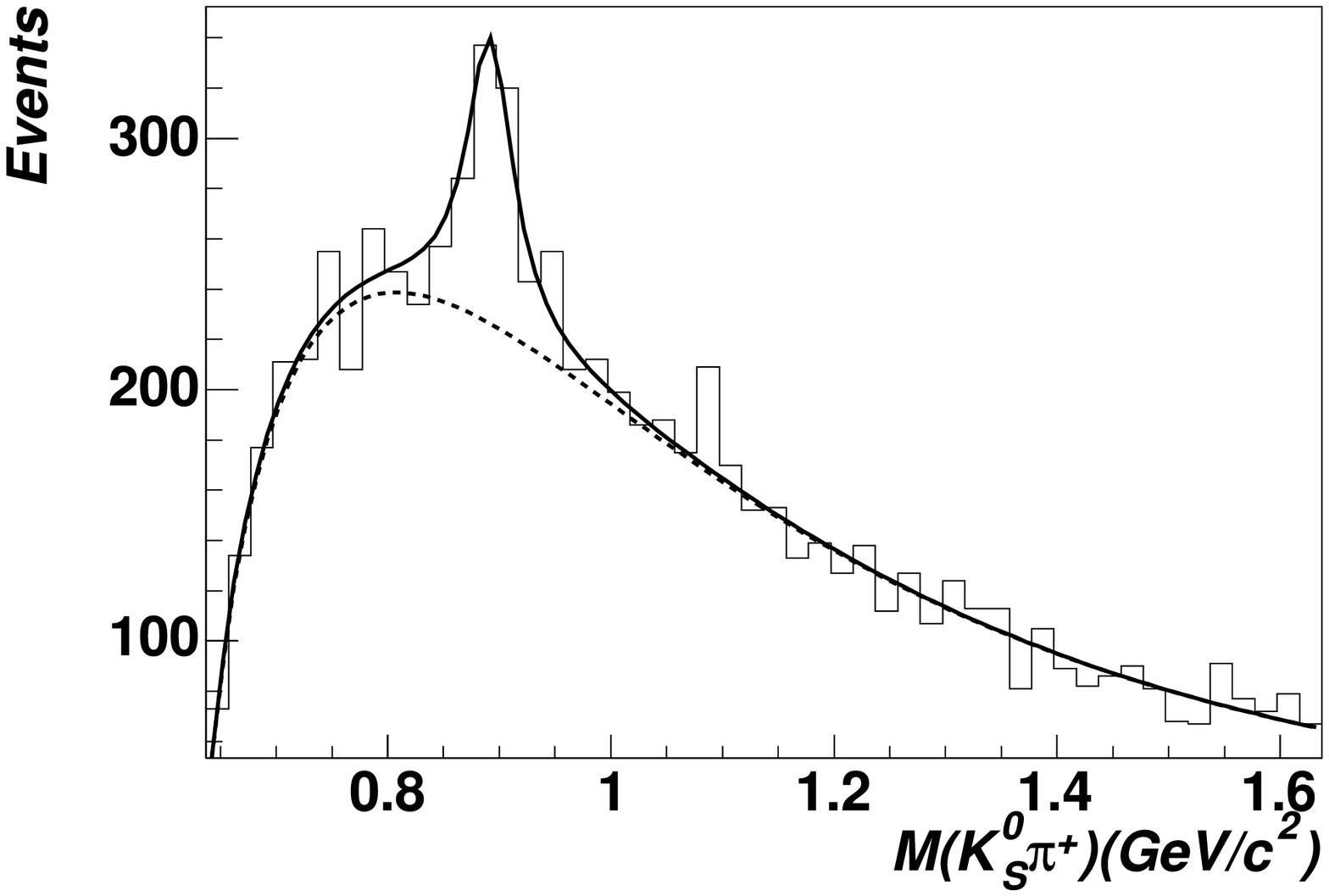,width=0.49\linewidth}}&
\mbox{\epsfig{file=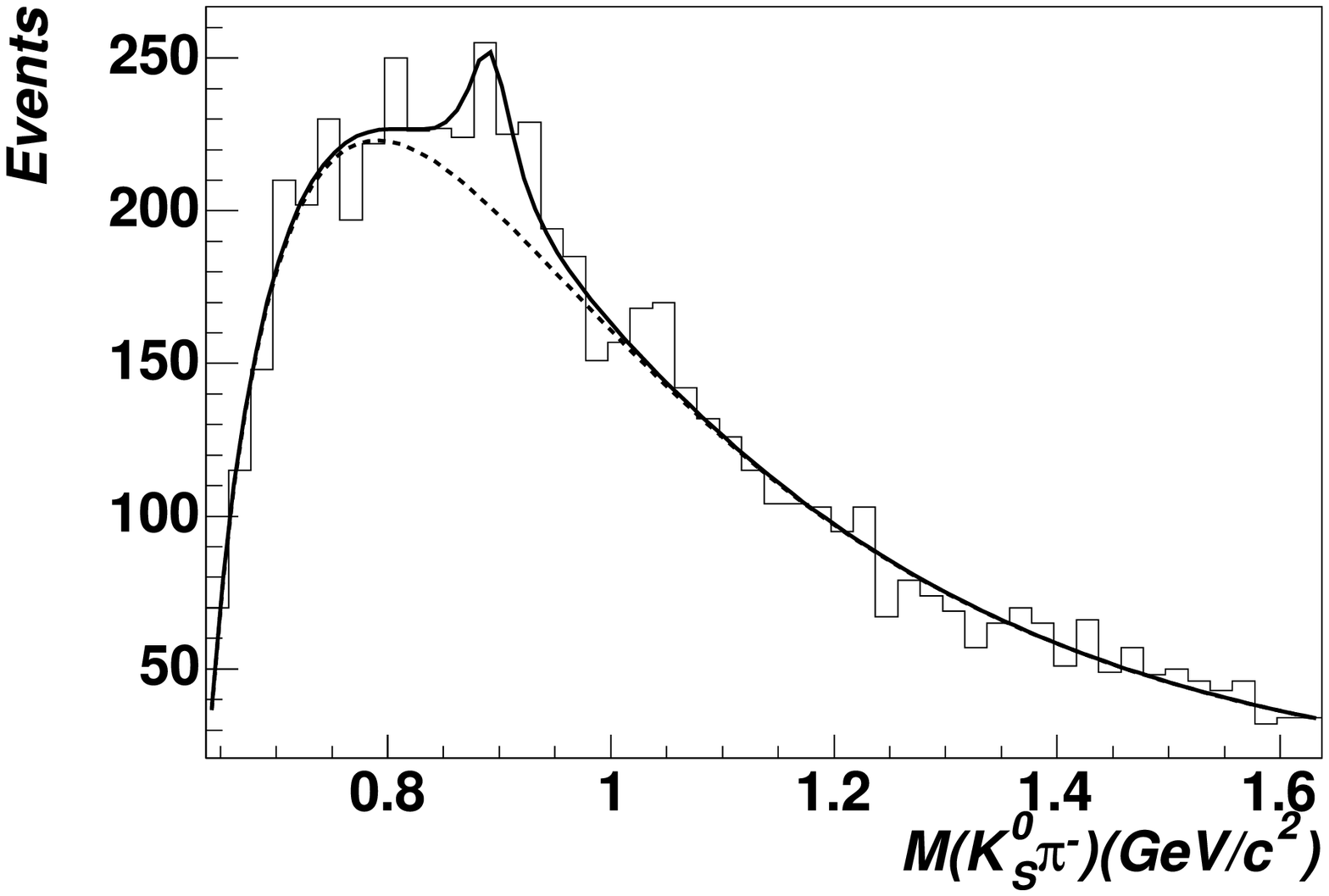,width=0.49\linewidth}}\\
\end{tabular}
\begin{picture}(0,0)
\put(-115,110){{\Large \bf ${K^\star}^+$}}
\put(95,110){{\Large \bf ${K^\star}^-$}}
\end{picture}
}
\protect\caption{\it $\rm \ko \pi^+$ (left) and $\rm \ko \pi^-$ (right) 
invariant mass distributions in the data.
}
\label{fig:kstar}
\end{figure}

\begin{figure}[htb]
\center{%
\begin{tabular}{cc}
\mbox{\epsfig{file=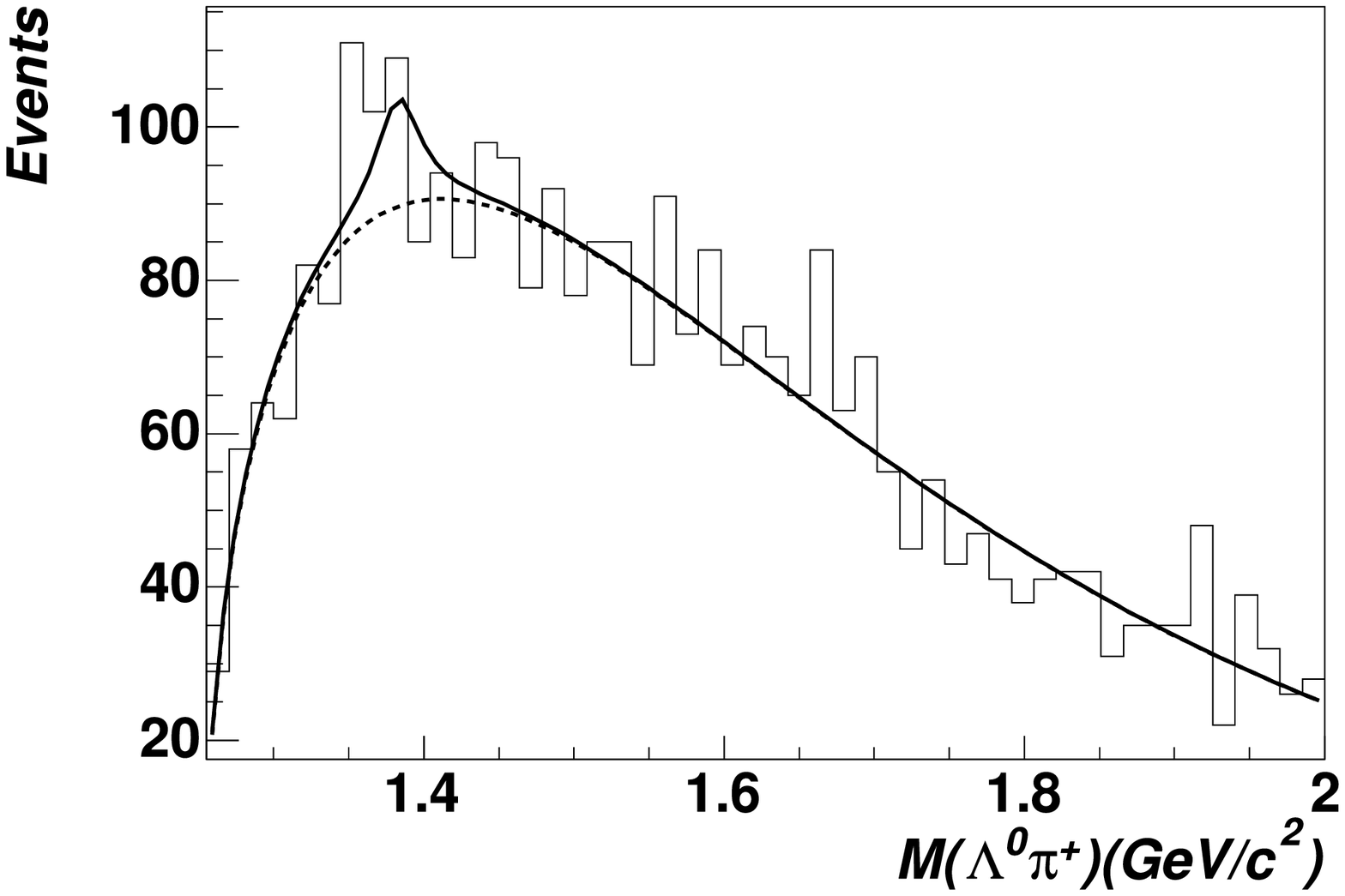,width=0.49\linewidth}}&
\mbox{\epsfig{file=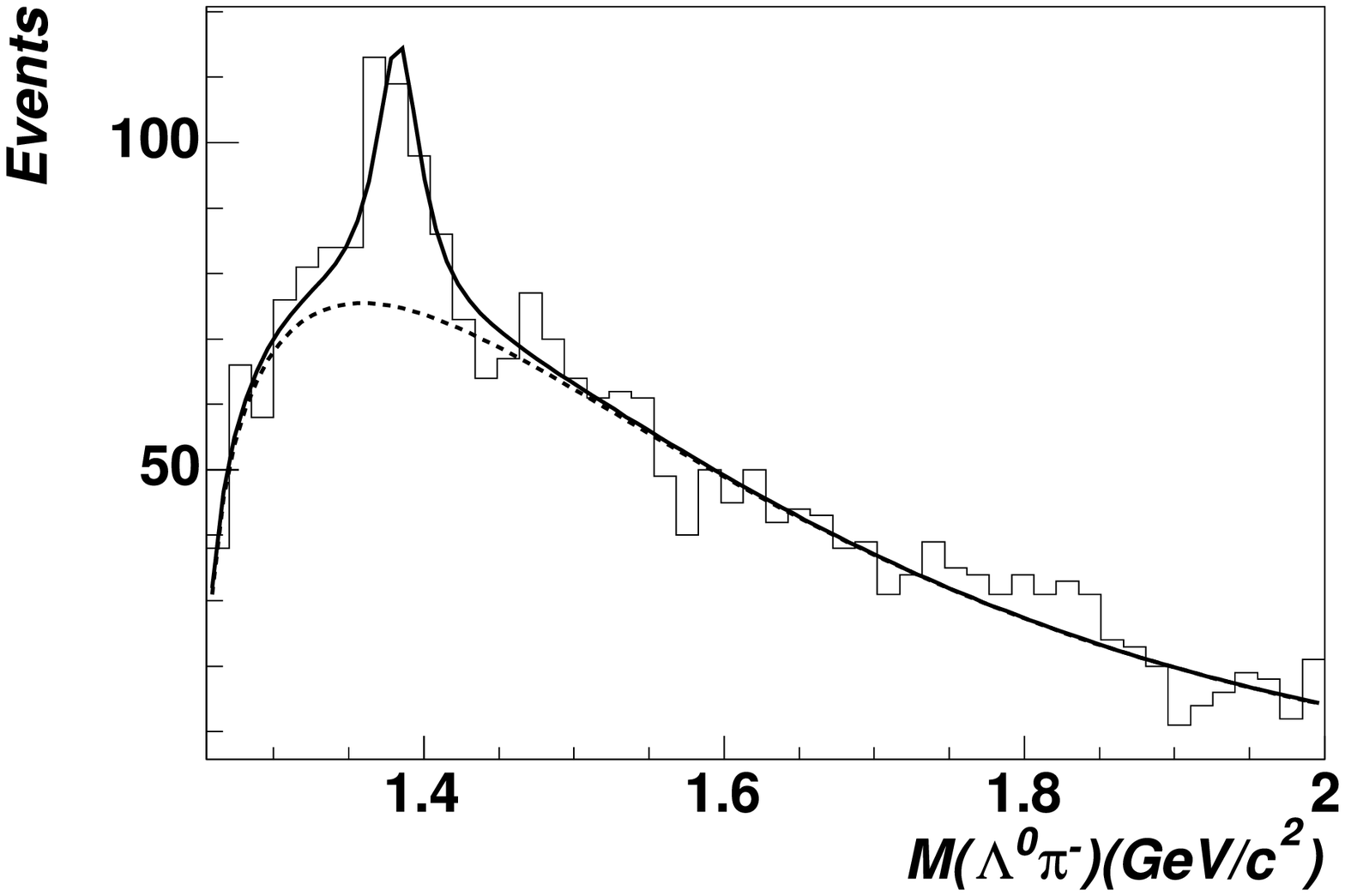,width=0.49\linewidth}}\\
\end{tabular}
\begin{picture}(0,0)
\put(-125,110){{\Large \bf $\rm {\Sigma(1385)}^+$}}
\put(80,110){{\Large \bf $\rm {\Sigma(1385)}^-$}}
\end{picture}
}
\protect\caption{\it $\rm \lam \pi^+$ (left) and $\rm \lam \pi^-$ (right) 
invariant mass distributions in the data.
}
\label{fig:sigmastar}
\end{figure}

We have studied the invariant mass distributions for ($\ko \pi^\pm$)
and ($\lam \pi^\pm$) combinations. The procedure used for the signal extraction is the 
one developed for the $\nu_\mu$ CC sample~\cite{nomad-strange-cc}.
Clear signals corresponding to 
${K^\star}^\pm$ (see Fig.~\ref{fig:kstar}) and ${\Sigma(1385)}^\pm$ (see
Fig.~\ref{fig:sigmastar}) have been observed. The results for the fraction of $\ko$
produced via ${K^\star}^\pm$ , $f_{K^\star} = ({K^\star}^\pm \rightarrow \ko \pi^\pm)/\ko$
are presented in Table~\ref{tab:k*_yields}. Similarly, the results for the fraction of
$\lam$ produced via ${\Sigma(1385)}^\pm$ decays,
$f_{\Sigma^\star} = ({\Sigma(1385)}^\pm \rightarrow {\lam} \pi^\pm)/{\lam}$
are presented in Table~\ref{tab:sigma*_yields}.
One can conclude that a reasonable agreement between data and MC
is obtained, in general, with the new set of JETSET parameters.

\begin{table}[htbp]
\begin{center}
\caption{ \it $\rm {K^\star}^{\pm} \to \ko \pi^\pm$ summary in
NC events. The fraction $f_{K^\star}$ is defined in the text. The errors are statistical only.
\label{tab:k*_yields}
}
\vspace*{0.3cm}
\begin{tabular}{||c|c|c|c||}
\hline
\hline
$K^\star$ type & Entries     & $f_{K^\star}$, (\%)& ${f_{K^\star}}_{DATA}/{f_{K^\star}}_{MC}$\\
\hline
\hline
$K^{\star +}$  &$526 \pm 72$ & $14 \pm 2$   & $1.1 \pm 0.2$ \\
$K^{\star -}$  &$246 \pm 67$ & $6 \pm 2$   & $0.7 \pm 0.2$ \\
\hline
\hline
\end{tabular}
\end{center}
\end{table}

%
%
\begin{table}[htbp]
\begin{center}
\caption{ \it $\rm {\Sigma(1385)}^{\pm} \to \lam \pi^\pm$ summary in
NC events. The fraction $f_{\Sigma^\star}$ is defined in the text. The errors are statistical only.}
\label{tab:sigma*_yields}
\vspace*{0.3cm}
\begin{tabular}{||c|c|c|c||}
\hline
\hline
$\Sigma(1385)$ type & Entries     & $f_{\Sigma^\star}$, (\%)& ${f_{\Sigma^\star}}_{DATA}/{f_{\Sigma^\star}}_{MC}$\\
\hline
\hline
$\Sigma(1385)^+$  &$47 \pm 34$ & $2 \pm 2$   & $0.5 \pm 0.3$ \\
$\Sigma(1385)^-$  &$125 \pm 32$ & $6 \pm 2$   & $1.4 \pm 0.4$ \\
\hline
\hline
\end{tabular}
\end{center}
\end{table}

\section{\label{sec:polar}Measurement of the $\lam$ polarization}

Measurement of the $\lam$ polarization in $\nu$ NC
interactions can provide some additional information with respect to 
the results obtained for $\nu_\mu$ CC events since NC
interactions are different at the quark level~\cite{Mangano,EKN}. 

The $\lam$ polarization is measured by the {\em asymmetry} 
in the angular distribution of the protons in the 
parity violating decay process $\lamdecay$.
In the $\Lambda$ rest frame the decay protons are distributed as:
\begin{equation}
\frac{1}{N}\frac{d N}{d \Omega} = \frac{1}{4\pi}(1+\alpha_\Lambda
\mathbf P \cdot \mathbf k),
\label{eq:asymmetry}
\end{equation}
where $\mathbf P$ is the $\lam$ polarization vector,
$\alpha_\Lambda = 0.642 \pm 0.013$~\cite{PDG} is the decay asymmetry parameter 
and $\mathbf k$ is the unit vector along the decay proton direction.

Since it is not possible to reconstruct the $Z^0$ direction in neutral current
events as distinct from 
the $W$ direction in the charged current events one has to redefine 
the coordinate system for the polarization measurement compared to the
CC case~\cite{nomad-lambda-polar}. 
For the polarization measurements in the NC sample 
we have chosen the coordinate system in the laboratory
rest frame which is Lorentz invariant with respect to a shift
along the $\vo$ velocity as follows:
\begin{itemize}
\item the $\bm n_x$ axis is chosen along the reconstructed 
$\vo$ direction ($\bm e_{V^0}$);
\item the $\bm n_y$ axis is defined as:\\
$\bm n_y= \bm e_{V^0} \times \bm e_\nu $, where $\bm e_\nu$ is the unit
vector along the incoming neutrino direction.
\item the $\bm n_z$ axis is chosen to form a right-handed coordinate system:
 \\
$\bm n_z = \bm n_x \times \bm n_y$. 
\end{itemize}

For the polarization measurements we have used the 3D method developed
for the CC sample and described in detail in~\cite{nomad-lambda-polar}.

The $\Lambda$ results are summarized in Table~\ref{tab:lam_polar_results}.
We observe a negative polarization along the $\vo$
direction and in the direction orthogonal to the $(\nu, \vo)$ plane.
Qualitatively these results are consistent with the longitudinal and
transverse $\lam$ polarization observed in $\nu_\mu$ CC 
interactions~\cite{nomad-lambda-polar}.
The
results for the $\alam$ case are inconclusive because of the large
statistical errors.  

\begin{table}[htbp]
\begin{center} 
\caption{\it The $\lam$ polarization in NC
  interactions. Both statistical and systematic errors are shown.}
\vspace*{0.3cm}
\label{tab:lam_polar_results}
\begin{tabular}{c|c|c}
\hline
\hline
$P_x$         &  $P_y$         &$P_z$ \\
\hline
$-0.23 \pm 0.08 \pm 0.05$ & $-0.19 \pm 0.07 \pm 0.04$& $0.01 \pm 0.07 \pm 0.03$\\
\hline
\hline
\end{tabular}
\end{center} 
\end{table}

To check the stability of the results we examined the following sources of
systematic uncertainties:
\begin{itemize}
\item[-] dependence of final results on the selection criteria of $V^0$'s 
         which include possible effects related to the contamination from 
         fake $\Lambda$ and secondary interactions. These selection criteria 
         were varied as described in Ref.~\cite{nomad-lambda-polar}. 
\item[-] dependence of final results on the neutral current selection criteria.
         The $\ln \lambda_{6}^{\rm NC}$ cut was varied in the interval from 0 to 1.
\end{itemize}
For each of these sources of uncertainties the maximum deviation from the 
reference result was found. These were then added in quadrature to give the 
total systematic uncertainty. 

As a cross-check we measured the polarization vector 
of $\ko$ mesons setting the decay asymmetry parameter $\alpha$ to 1
and found no significant fake asymmetry 
(see Table~\ref{tab:k0_polar_results}).

\begin{table}[htbp]
\begin{center} 
\caption{\it The $\ko$ ``polarization'' in NC interactions. 
Both statistical and systematic errors are shown.
\label{tab:k0_polar_results}
}
\vspace*{0.3cm}
\begin{tabular}{c|c|c}
\hline
\hline
$P_x$         &  $P_y$         &$P_z$ \\
\hline
$-0.03 \pm 0.03 \pm 0.01$& $-0.01 \pm 0.03 \pm 0.01$& $-0.01 \pm 0.03 \pm 0.03$\\
\hline
\hline
\end{tabular}
\end{center} 
\end{table}

The results on the $\lam$ polarization are compared to 
the phenomenological predictions within the framework of the intrinsic polarized
strangeness model~\cite{EKN}. 
This is the only model currently available which can be applied to the
whole $x_F$ range. Other models~\cite{Mangano,Anselmino:2001ey} 
are valid only in the $x_F > 0$ interval.

\begin{figure}[htb]
  \centering\epsfig{file=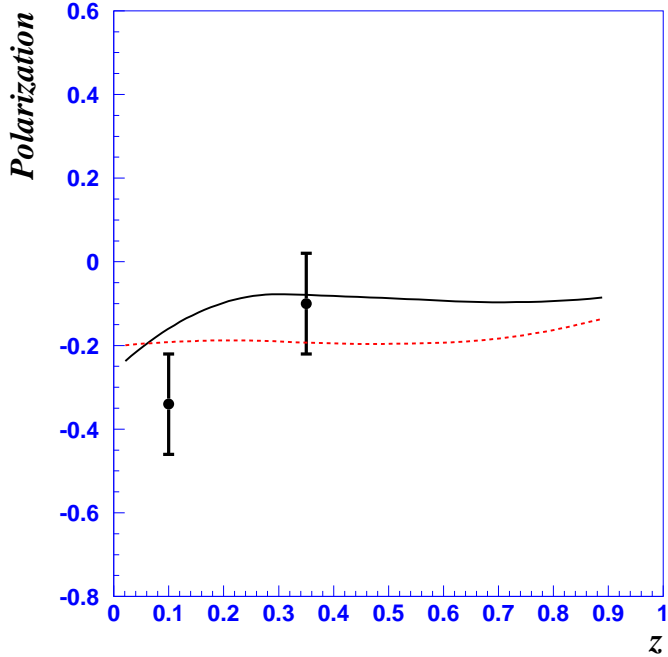,width=0.7\textwidth} 
  \caption{\label{fig:numuxnc_z} The $\lam$ polarization as a function
  of $z$ measured in NOMAD compared to the intrinsic polarized strangeness
  model~\cite{EKN} predictions. Solid and dashed curves correspond to two 
  different schemes considered in the paper~\cite{EKN} and named A and B,
  respectively (see text).}  
\end{figure}

In Fig.~\ref{fig:numuxnc_z} we display the $z$
dependence of the longitudinal $\lam$ polarization measured by NOMAD 
and predicted by the model~\cite{EKN} (with the cut $E_{jet}>3$ GeV). 
While the coordinate system of the model~\cite{EKN} is slightly different from that 
chosen by us (the x-axis in~\cite{EKN} is along the $Z$ boson direction) we expect 
the angle between these two coordinate systems to be rather small. 

The predictions are made for two extreme schemes: 
in the first scheme (model A) the $\Lambda$ hyperon contains either 
the outgoing quark or the spectator diquark system; in the second scheme 
(model B) the $\Lambda$ hyperon is associated with the fragmentation 
of either the outgoing quark or the spectator diquark depending on 
their phase space distance to the $\Lambda$. One can conclude that 
within statistical errors both models are in reasonable agreement 
with our data. It is important to note that the model is tuned to the 
$\lam$ polarization measurements performed by NOMAD in the $\nu_\mu$ CC
sample, which are sensitive only to the $d-s$ quark spin
correlations. The measurements from NC interactions suggest that 
the $d-s$ and $u-s$ spin correlations are similar.  

\section{\label{sec:conclusions}Conclusion}

In this article we presented the results of our study of strange hadrons produced 
in neutrino neutral current interactions using the data from the 
NOMAD experiment. We applied an algorithm based on the event kinematics 
to reduce the charged current contamination in the NC sample. The background in the final sample is estimated to be about 8\% of which 6\% correspond to $\nu_\mu$ CC 
contribution and the remaining 2\% originate from $\bar\nu_\mu, \nu_e, \bar\nu_e$ CC contributions. 
 
To identify neutral strange hadrons we applied the $\vo$ identification procedure based 
on a kinematic fit. As a result we found very clean $\vo$ samples with a  
background contamination of the same order as in $\nu_\mu$ CC data. 
In 226681 identified NC interactions we have observed 3691 $\ko$'s,
1619 $\lam$'s and 146 $\alam$'s about 60 times more than
previously published results~\cite{E632}. 
Integral yields of neutral strange particles 
have been measured to be: $(5.16 \pm 0.14 (stat.) \pm 0.09 (syst.))\%$, 
$(0.43 \pm 0.04 (stat.)\pm 0.03 (syst.))\%$, 
$(8.62 \pm 0.15 (stat.)\pm 0.11 (syst.))\%$ for $\lam$, $\alam$ and $\ko$ respectively. 
These yields agree within few \% with those measured in CC interactions. 
As observed by the E632 collaboration~\cite{E632} at higher average
neutrino energies the yields of both $\ko$ and $\lam$ are comparable
in NC and CC interactions.

Reasonable agreement has been obtained for the $z$ distribution of neutral
strange particles between the data and the predictions from a
MC simulation tuned on the CC sample.

Decays of resonances and heavy hyperons with identified $\ko$ and $\lam$ in 
the final state have been analyzed as well. Clear signals corresponding to 
$\rm {K^\star}^\pm$ and $\rm {\Sigma(1385)}^\pm$ have been observed. 

First results on the measurements of the $\lam$ 
polarization in neutrino neutral current interactions are
presented. 
The results are close to those observed in $\nu_\mu$ CC interactions with a somewhat larger statistical uncertainty. The results for the longitudinal polarization of $\lam$ hyperons are in agreement with the predictions of the intrinsic polarized strangeness model~\cite{EKN}.

{\large \bf Acknowledgements}

We gratefully acknowledge  the CERN SPS accelerator and beam-line staff
for the magnificent performance of the neutrino beam.\
The experiment was  supported by  the following
funding agencies:
Australian Research Council (ARC) and Department of Education, Science and
Training (DEST), Australia;
Institut National de Physique Nucl\'eaire et Physique des Particules (IN2P3), 
Commissariat \`a l'Energie Atomique (CEA),  France;
Bundesministerium f\"ur Bildung und Forschung (BMBF, contract 05 6DO52), 
Germany; 
Istituto Nazionale di Fisica Nucleare (INFN), Italy;
Joint Institute for Nuclear Research and 
Institute for Nuclear Research of the Russian Academy of Sciences, Russia; 
Fonds National Suisse de la Recherche Scientifique, Switzerland;
Department of Energy, National Science Foundation (grant PHY-9526278), 
the Sloan and the Cottrell Foundations, USA.

\end{document}